\documentclass{aa}
\usepackage[varg]{txfonts}
\usepackage{graphicx}
\usepackage{natbib}
\bibpunct{(}{)}{;}{a}{}{,} 

\usepackage{hyperref}
\hypersetup{
    colorlinks = true,
    citecolor ={blue}
}

\usepackage{tabularx} 
\newcommand\clearrow{\global\let\rowmac\relax} 
\clearrow
\usepackage[normalem]{ulem}

\begin{document}

\title{Pulsating hydrogen-deficient white dwarfs and pre-white dwarfs observed with {\it TESS}}
\subtitle{II. Discovery of two new GW Vir stars: TIC\,333432673 and TIC\,095332541}

\author{Murat Uzundag\inst{1,2},
        Alejandro H. C\'orsico\inst{3,4}, 
        S. O. Kepler\inst{5}, 
        Leandro G. Althaus\inst{3,4},
        Klaus Werner\inst{6},
        Nicole Reindl\inst{7},
        Keaton J. Bell\inst{8,9},
        Michael Higgins\inst{10},
        Gabriela O. da Rosa\inst{5},
        Maja Vu\v{c}kovi\'{c}\inst{1}, 
        Alina Istrate\inst{11}
        }
           \institute{Instituto de F\'isica y Astronom\'ia, Universidad de Valpara\'iso, Gran Breta\~na 1111, Playa Ancha, Valpara\'iso 2360102, Chile
           \\
           \email{murat.uzundag@postgrado.uv.cl}
           \and
           European Southern Observatory, Alonso de Cordova 3107, Santiago, Chile
           \and
           Grupo de Evoluci\'on Estelar y Pulsaciones. 
           Facultad de Ciencias Astron\'omicas y Geof\'{\i}sicas, 
           Universidad Nacional de La Plata, 
           Paseo del Bosque s/n, 1900 
           La Plata, 
           Argentina
           \and
           IALP - CONICET
           \and
           Instituto de F\'{i}sica, Universidade Federal do Rio Grande do Sul, 91501-970, Porto-Alegre, RS, Brazil
           \and
           Institut f\"ur Astronomie und Astrophysik, Kepler Center for Astro and Particle Physics, Eberhard Karls Universit\"at, Sand 1,72076 T\"ubingen, Germany
           \and
           Institute for Physics and Astronomy, University of Potsdam, Karl-Liebknecht-Str. 24/25, D-14476 Potsdam, Germany
           \and
           DIRAC Institute, Department of Astronomy, University of Washington, Seattle, WA-98195, USA
           \and 
           NSF Astronomy and Astrophysics Postdoctoral Fellow
           \and
           Department of Physics, Duke University, Durham, NC-27708, USA      
           \and
           Department of Astrophysics/IMAPP, Radboud University, P O Box 9010, NL-6500 GL Nijmegen, The Netherlands
\\
        }
\date{}

  \abstract{
The {\it TESS}  mission is revolutionizing the blossoming area of asteroseismology, 
particularly of pulsating white dwarfs and pre-white dwarfs, thus 
continuing the impulse of its predecessor, the {\it Kepler} mission.}
{In this paper,  we present the observations from the extended \textit{TESS} mission in both 120\,s short-cadence and 20\,s ultra-short-cadence mode of two pre-white dwarf stars showing hydrogen deficiency. 
We identify them as two new GW Vir stars, TIC\,333432673 and TIC\,095332541. 
We apply the tools of asteroseismology with the aim of deriving their structural 
parameters and seismological distances.}
{We carried out a spectroscopic analysis and a spectral fitting of TIC\,333432673 and TIC\,095332541.  We also processed and analyzed the high-precision {\it TESS} 
photometric light curves of the two target stars, and derived 
their oscillation frequencies. We performed an asteroseismological analysis of 
these stars on the basis of  PG~1159 evolutionary models that take into 
account the  complete evolution of the progenitor stars. We searched for 
patterns of uniform period spacings in order to constrain the  stellar  
mass of the stars, and employed the individual observed periods to search for a 
representative seismological model.}
{The analysis of the {\it TESS} light curves of TIC\,333432673 and TIC\,095332541 reveals the presence of several oscillations with periods ranging from 350 to 500~s associated to typical gravity ($g$)-modes. From follow-up ground-based spectroscopy, we find that both stars have similar effective temperature ($T_\mathrm{eff} = 120,000 \pm 10,000$\,K) and surface gravity ($\log g = 7.5 \pm 0.5$) but a different He/C composition of their atmosphere. On the basis of PG~1159 evolutionary tracks,
we derived a spectroscopic mass of $M_{\star}$ = $0.58^{+0.16}_{-0.08}\,M_{\odot}$ 
for both stars. Our asteroseismological analysis of TIC\,333432673
allowed us to find a constant period spacing compatible with a stellar mass  
$M_{\star}\sim 0.60-0.61\,M_{\odot}$, and an asteroseismological model for this
star with a stellar mass $M_{\star}$ = $0.589\pm 0.020$ $M_{\odot}$, and a seismological 
distance of $d= 459^{+188}_{-156}$ pc. 
For this star, we find an excellent agreement between the  different methods to infer the stellar mass, and also between the seismological distance and that measured with {\it Gaia} 
($d_{\rm Gaia}= 389^{+5.6}_{-5.2}$ pc). 
For TIC\,095332541, we have found a possible period spacing that suggests
a stellar mass of $M_{\star}\sim 0.55-0.57\,M_{\odot}$. Unfortunately, we have not been able to find an asteroseismological model for this star.}
{Using the high-quality data collected by the {\it TESS} space mission and follow-up spectroscopy, 
we have been able to discover and characterize two new GW Vir stars. The {\it TESS} mission 
is having, and will continue to have, an unprecedented impact on the area of white-dwarf asteroseismology.}

\keywords{asteroseismology --- stars: oscillations (including pulsations) --- stars: interiors --- 
stars:  evolution --- stars: white dwarfs}
\authorrunning{C\'orsico et al.}
\titlerunning{Asteroseismology of GW Vir stars with {\it TESS}}
\maketitle


\section{Introduction}

GW Vir stars are pulsating PG~1159 stars, that is, pulsating hot hydrogen (H)-deficient,
carbon (C)-,~ oxygen (O)-, helium (He)-rich white dwarf (WD) and pre-WD stars.
PG~1159 stars represent the evolutionary link between post-AGB stars and most of the
H-deficient WDs, including DO and DB WDs \citep{2006PASP..118..183W}.
These stars likely have their origin in a born-again episode induced
by a post-AGB He thermal pulse \citep[see][for references]{2001ApJ...554L..71H,2001Ap&SS.275....1B,
  2005A&A...435..631A,2006A&A...449..313M}. GW~Vir stars constitute 
  the hottest class of pulsating WDs and pre-WDs, the other categories 
  being the DAV or ZZ Ceti (H-rich atmospheres) stars, DBV or V777 Her 
  (He-rich atmospheres) stars, ELMV stars (H-rich atmospheres and extremely low masses), 
  and pre-ELMV stars, likely the precursors of the ELMVs  \citep[see the reviews by][]
  {2008ARA&A..46..157W, 2010A&ARv..18..471A,2019A&ARv..27....7C}.
  The category of GW Vir stars 
  includes PNNV stars, which are still surrounded by a nebula, and  DOV stars, 
  that lack a nebula \citep{1991ApJ...378..326W}, and also  the pulsating Wolf-Rayet
central stars of planetary nebula ([WC] stars)  and Early-[WC] = [WCE] stars, because they 
share the same pulsation properties of pulsating PG~1159 stars \citep{2007ApJS..171..219Q}.  
GW Vir stars exhibit multiperiodic luminosity variations with periods in 
the range 300--6000 sec, originated from $g$(gravity)-mode pulsations 
excited by the $\kappa$-mechanism due to partial ionization of 
C and O in the outer layers\footnote{Due to the high surface temperatures of 
these stars, it is likely that they do not have convection in their envelopes, 
making them one of the few classes of pulsating stars for which the
well known complication of the convection-pulsation interaction in the pulsational 
stability analyses is not present.} \citep{1983ApJ...268L..27S,1984ApJ...281..800S,
  1991ApJ...383..766S,1996ASPC...96..361S,1997A&A...320..811G, 2005A&A...438.1013G, 2006A&A...458..259C,2007ApJS..171..219Q}.   

Asteroseismology of WDs and pre-WDs has been strongly promoted during the 
last decade mainly by the availability of space missions that provide unprecedented high-quality data. Particularly, the {\it Kepler} satellite, 
both the main mission \citep{2010Sci...327..977B} and the {\it K2} mode \citep{2014PASP..126..398H}, allowed the study of 32 ZZ Ceti stars and two V777 Her stars \citep{2011ApJ...736L..39O,2014MNRAS.438.3086G, 2014ApJ...789...85H, Bell_2015, 2017ApJS..232...23H,2017ApJ...835..277H,
  2017ApJ...851...24B,2020FrASS...7...47C}, until
it was out of operation by October 2018.  The successor of {\it
Kepler} is the Transiting Exoplanet Survey Satellite \citep[{\it
TESS},][]{2015JATIS...1a4003R}.  {\it TESS} has provided extensive 
photometric observations of the $200\,000$ brightest stars in 85 \% 
of the sky in the first part of the mission, each observation with a 
time base of about 27 days per sector observed. In the context 
of pulsating WDs and pre-WDs, the high-quality observations of {\it TESS}, 
combined with ground-based observations, are
able to provide a  very important input to the asteroseismology of DBVs
\citep[][]{2019A&A...632A..42B,2021arXiv210317192B},   
DAVs  \citep[][]{2020A&A...638A..82B},  pre-ELMVs 
\citep{2020ApJ...888...49W,2021AJ....161..137H}, and GW
Vir stars \citep{2021A&A...645A.117C}. 

In this work, we report for the first time the photometric variability of the 
PG1159 stars TIC\,333432673 (WD~J064115.64-134123.77) and TIC\,095332541 (WD~J060244.99-135103.57) observed with {\it TESS}. 
Given the small number of already known pulsating  stars of this class \citep[20 objects; see][]{2019A&ARv..27....7C,2021arXiv210808167S}, the discovery of two new GW Vir stars constitutes a particularly relevant finding, even more since the {\it Kepler}/{\it K2} mission publications do not include any new object of this nature to date. We perform an asteroseismological analysis of these stars based on the state-of-the-art evolutionary models of PG~1159 stars of \cite{2005A&A...435..631A} and  \cite{2006A&A...454..845M}.   
This study is the second part of our series of papers devoted to the study of pulsating H-deficient WDs observed with {\it TESS}. The first article was devoted to a set of six 
already known GW Vir stars \citep{2021A&A...645A.117C}.

The paper is organized as  follows. In  Sect. \ref{spectroscopic observations}, we 
present the details of the spectroscopic observations and the data reduction. 
In  Sect. \ref{spectral fit}, we derive atmospheric parameters for each star 
by fitting synthetic spectra to the newly obtained low-resolution spectra.
In  Sect. \ref{photometric_observations}, we analyse the photometric \textit{TESS} data and give details on the frequency analysis. 
Sect. \ref{ast_analysis} is devoted to the asteroseismic analysis of our targets.
Finally, in Sect. \ref{conclusions}, we  summarize  our  main  results.

\section{Spectroscopy}
\label{spectroscopic observations}  

TIC\,333432673 (WD\,J064115.64-134123.77) and TIC\,095332541 (WD\,J060244.99-135103.57) were classified as white dwarf candidates by \citet{GentileFusillo19} from their colors and \textit{Gaia} DR2 parallax.
The {\it Gaia} DR3 parallax and corresponding distance for TIC\,333432673 are  $\pi= 2.57^{+0.07}_{-0.04}$~mas and $d = 389.00^{+5.59}_{-5.22}$~pc, while for TIC\,095332541 are  $\pi= 2.60^{+0.07}_{-0.04}$~mas and $d = 384.48^{+5.54}_{-5.04}$~pc \citep{Bailer-Jones2021}, respectively.
We obtained spectroscopic observations for TIC\,333432673 and TIC\,095332541, to determine the atmospheric parameters. 

TIC\,333432673 was observed with the Southern Astrophysical
Research (SOAR) Telescope, a 4.1-meter aperture optical and near-infrared telescope \citep{clemens2004}, situated at Cerro Pach\'on, Chile on March 5, 2021 (under the program allocated by the Chilean Time Allocation Committee (CNTAC), no:CN2020B-74).
We use Goodman spectrograph with a setup of 400\,l/mm grating with the blaze wavelength 5500 $\AA$  (M1: 3000-7050 \AA) with a slit of 1 arcsec. This setup provides a resolution of about 5.6~\AA.
The data reduction has been partially done by using the instrument pipeline\footnote{\url{https://github.com/soar-telescope/goodman_pipeline}} including overscan, trim, slit trim, bias and flat corrections. 
For cosmic rays identification and removal, we used an algorithm as described by \citet{wojtek2004}, which is embedded in the pipeline.
Then, we applied the wavelength calibrations using the frames obtained with the internal He-Ar-Ne comparison lamp at the same telescope position as the targets.
In a last step, we normalized the spectrum with a high-order Legendre function, using the standard star EG\,274 observed with the same setup. 
The signal to noise ratio of the final spectrum is around 80 at 4200 $\AA$ with 450\,s exposure time.

TIC\,095332541 was observed with the Isaac Newton Telescope (INT), a 2.54~m (100~in) optical telescope run by the Isaac Newton Group of Telescopes at Roque de los Muchachos Observatory on La Palma on February 15 and 16, 2021 (ProgID: ING.NL.21A.003) with an exposure time of $1800$\,s, respectively. We used the Intermediate Dispersion Spectrograph (IDS) long-slit spectrograph with the grating R400V ($R=1452$) and a 1.5 arcsec slit. This setup provides a resolution of about 3.5~\AA.
Bias and flat field corrections were applied to the data, the wavelength calibration was performed with Cu-Ne-Ar calibration lamp spectra. We did not flux calibrate the spectra.
The signal-to-noise ratio (SNR) of the final spectra is around 80 (see Table \ref{tablespectroscopy}).

\begin{table*}
\renewcommand{\arraystretch}{1.2}
\caption{Log of spectroscopic observations.}
\begin{tabular}{ccccccc}
\hline \hline
TIC & Name & RA & Dec  & Exp.  & S/N & Telescope/Inst. \\
    &      & (J2000)  &  (J2000) & (sec) & (at 4200 \AA)  &   \\
\hline                                           
333432673 & WD~J064115.64$-$134123.77 & 06:41:15.88 & -13:41:31.31  & 450  &  80  & SOAR/GOODMAN \\
095332541 & WD~J060244.99$-$135103.57 & 06:02:45.00 & -13:51:03.50  & 2x1800 &  80  & INT/IDS    \\

\hline 
\label{tablespectroscopy}
\end{tabular}
\end{table*}

\section{Spectral fitting}
\label{spectral fit}

The spectra of both stars exhibit lines exclusively from He\,II and C\,IV. Oxygen, which is usually the most abundant element after He and C in PG1159 stars, and nitrogen, which is found as trace element in some PG1159 stars, might be detectable in spectra of better resolution and signal-to-noise ratio. There are no hints of the presence of hydrogen in the spectra.

For the spectral analysis, we used a grid of line-blanketed non-local thermodynamic equilibrium (non-LTE) model atmospheres consisting of H, He, and C as introduced by \citet{2014A&A...564A..53W}. In essence, it spans $T_\mathrm{eff}$ = 60,000--140,000\,K in effective temperature and $\log g$ = 4.8--8.3 in surface gravity, with steps of 5,000\,K or 10,000\,K and 0.3\,dex, respectively. C/He mass ratios in the range 0.0--1.0 were considered, namely C/He = 0.0, 0.03, 0.09, 0.33, 0.77, and 1.0.  Synthetic spectra were convolved with a Gaussian accounting for the spectral resolution of the observations. The best fitting models were chosen by visual comparison with the rectified observed spectra.

\begin{figure*}[t]
 \centering  \includegraphics[width=0.9\textwidth]{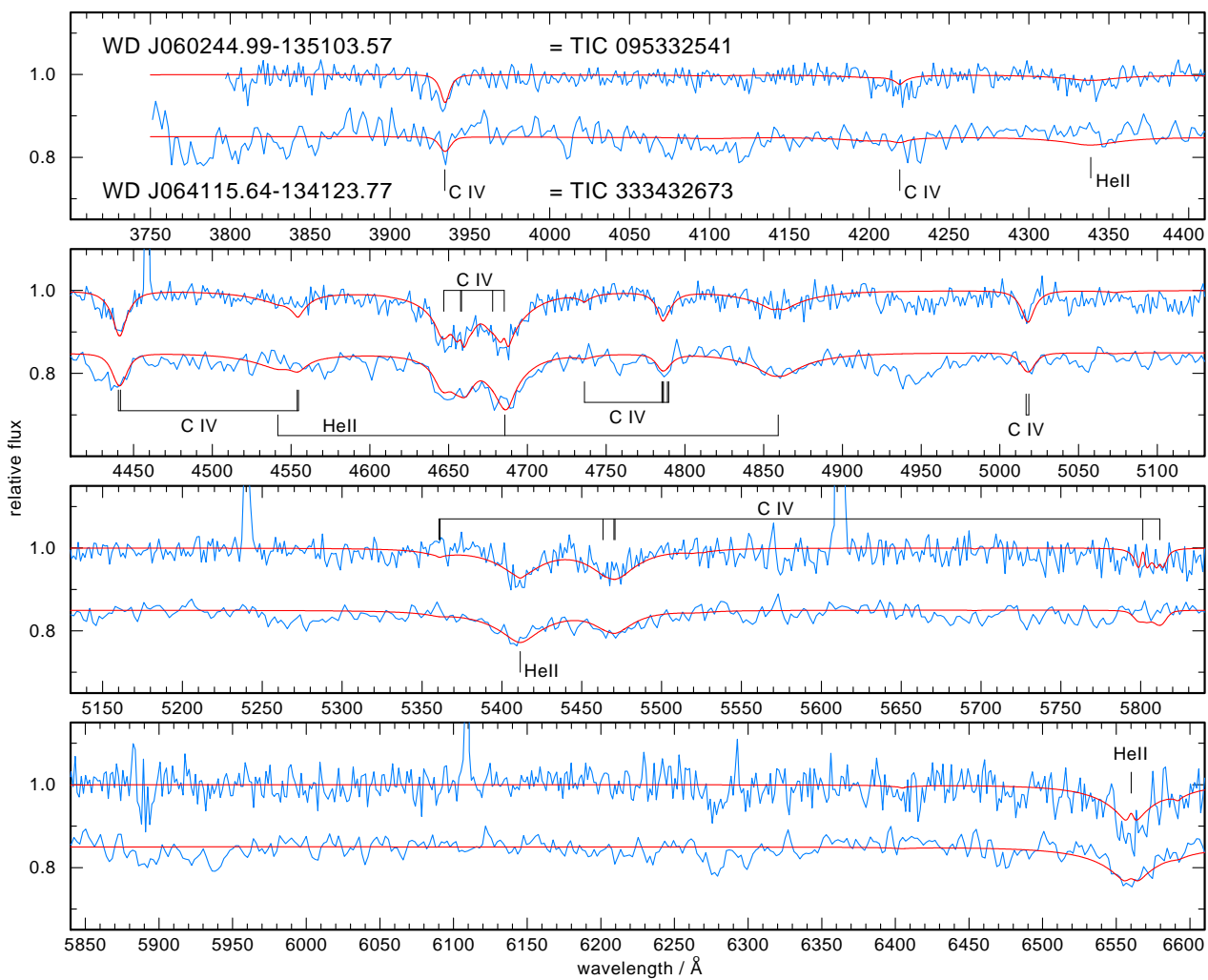}
  \caption{Optical spectra of the two new GW Vir stars. Overplotted are the best-fit models. Identifications of He\,II and C\,IV lines are marked.}
\label{fig:spectral_fits}
\end{figure*}

The model fits are displayed in Fig.\,\ref{fig:spectral_fits}. Both stars have $T_\mathrm{eff} = 120,000 \pm 10,000$\,K and $\log g = 7.5\pm0.5$, but a different atmospheric composition. For  TIC\,095332541, we found He = $0.50^{+0.20}_{-0.05}$  and C = $0.50^{+0.05}_{-0.20}$ and for TIC\,333432673, we measured He = $0.75^{+0.05}_{-0.15}$ and C = $0.25^{+0.15}_{-0.05}$ (mass fractions). 

\section{Photometric observations --- \textit{TESS}}
\label{photometric_observations}  

We investigate for variability of the two new
PG~1159 stars by examining their high-precision photometric observations
obtained with \textit{TESS}. TIC\,333432673 was observed with 120\,s cadence mode, while TIC\,095332541 was observed with both 120\,s and 20\,s cadence modes in Sector 33, between 2020-Dec-17 and 2021-Jan-13.
We downloaded the data from the “Barbara A. Mikulski Archive for Space Telescopes” (MAST)\footnote{\url{http://archive.stsci.edu}}. 
We downloaded the target pixel files (TPFs) of both targets from the MAST archive with the Python package $\tt{lightkurve}$ \citep{lightkurve2020}. 
The TPFs are examined to determine the amount of crowding and other potential bright sources near the target. 
The contamination factor is indicated with the keyword $\tt{CROWDSAP}$, which gives the ratio of the target flux to the total flux in the \textit{TESS} aperture. 
For each target, we have checked the contamination by looking at the $\tt{CROWDSAP}$ parameter which is listed in Table \ref{DOVlist}. 
In the case of TIC\,095332541 the $\tt{CROWDSAP}$ value is about 0.8, which implies that $\sim$ 20\% of the total flux measured in the \textit{TESS} aperture comes from other unresolved sources. We have checked for the nearby targets and their brightness. We found 3 nearby objects to TIC\,095332541 within 25 arcsec (based on $Gaia$ measurements) with $G_{mag}$ of 19.5, 20.78 and 20.76.
As the differences in magnitude between TIC\,095332541 and the 
nearby objects are larger than 4, we safely confirm that the variation comes from the PG~1159 star. In addition, these nearby objects are beyond the detection limit of \textit{TESS}. 



\begin{figure} 
\includegraphics[clip,width=1.\linewidth]{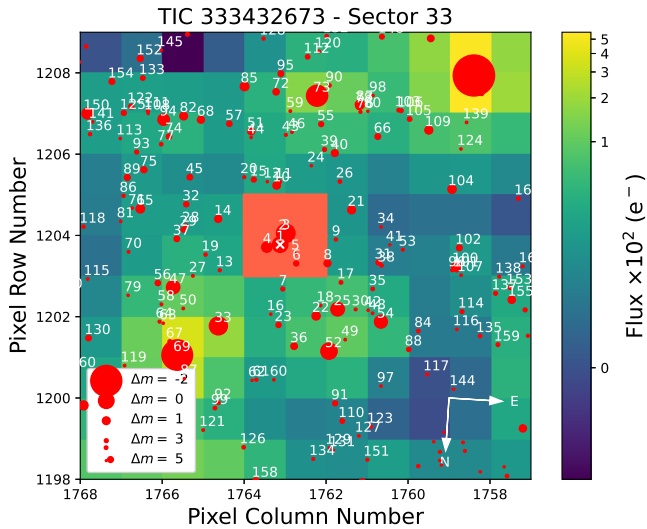}
\caption{Target pixel file (TPF) of TIC\,333432673 (created with $\tt{tpfplotter}$
, \citealt{Aller2020}).
The aperture mask used by the pipeline to extract the photometry is overplotted with 
red square. 
The size of the red circles indicates the Gaia magnitudes of all nearby stars and TIC\,333432673 (circle 1 is marked with a cross).}
\label{fig:FOV33} 
\end{figure}

In the case of TIC\,333432673, the $\tt{CROWDSAP}$ value is around 0.15, which means that more than 80\% of the flux in the \textit{TESS} aperture comes from the other blended sources. As the contamination from the nearby targets is non-negligible, we considered the proximity and brightness of nearby \textit{Gaia} sources that could contaminate the photometric aperture. 
In Fig. \ref{fig:FOV33}, we show the field of view of \textit{TESS} for TIC\,333432673 with the aperture mask. 
Within the aperture mask that we used to extract the photometry, four other objects along with TIC\,333432673 are located. Two of them are fainter than $G_{mag}= 18$ (circles 2 and 5), and thus it is not possible to detect any variation from them
with \textit{TESS}. 
However, the other two targets (circles 3 and 4) are relatively bright, with $G_{mag}$ of 14.7 and 16.4.
The origin of the signals has been found using a locator code (Higgins \& Bell, in prep.) that produces a light curve and computes its Fourier transform for each pixel in the star's Target Pixel File (TPF). Then it compares the amplitudes and locates the pixel where each signal period has the highest amplitude.
We find that the location of maximum power from the four highest signals in Table~\ref{table:333432673} is most consistent with the position of the PG\,1159 source (\textit{Gaia} \verb+source_id+ 2950907725113997312). 


The data are in the FITS format which includes all the photometric information, which have been already processed with the Pre-Search Data Conditioning Pipeline \citep{jenkins2016} to remove common instrumental trends. From the FITS file, first we have extracted times in barycentric corrected Julian days (“BJD - 245700”), and fluxes (“PDCSAP FLUX”). Afterwards, we removed outliers by applying a running $5\sigma$ clipping mask.
The fluxes were then normalized by the mean flux. 

\begin{figure} 
\includegraphics[clip,width=1.\linewidth]{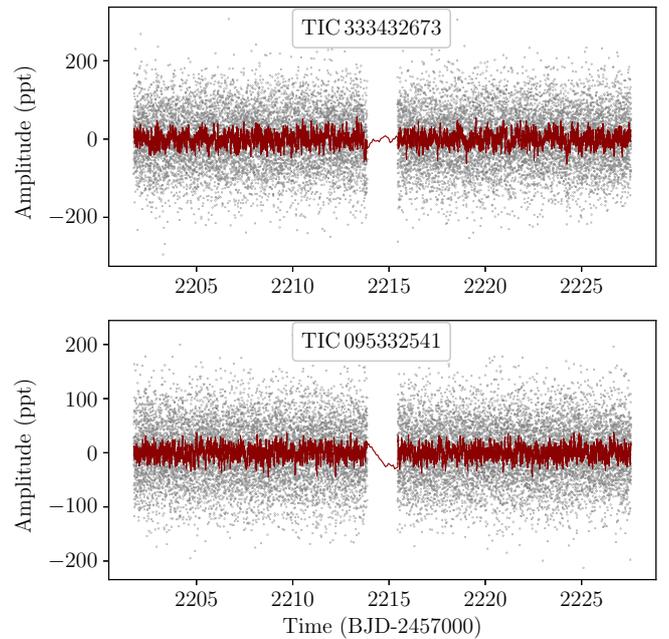}
\caption{The light curves of the new pulsating DOV stars TIC\,333432673 (upper panel) and TIC\,095332541 (lower panel).
The red lines are binned light curves which are calculated by running mean every 20 points (corresponding to 38 minutes).}
\label{fig:lightcurves} 
\end{figure}

\begin{figure} 
\includegraphics[clip,width=1.\linewidth]{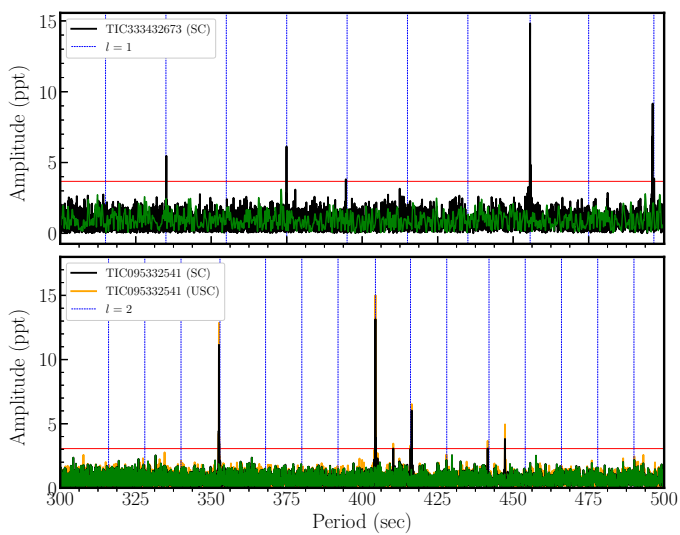}
\caption{
{ Pulsation spectrum in period of the new pulsating DOV stars TIC\,333432673 (upper panel) and TIC\,095332541 (lower panel). For TIC\,095332541, we overplotted the Fourier Transform of both 120~s data (black) and 20~s (orange). 
The horizontal red line indicates the 0.1\% false-alarm probability (FAP) level. In green we depict the FT of the prewhitened light curve.
The blue vertical dotted lines indicate the expected locations of $\ell=1$ modes for TIC\,333432673 and $\ell=2$ modes for TIC\,095332541 from the asymptotic pulsation theory described in Section \ref{mean_period_spacing-TIC333}.}}
\label{fig:FT} 
\end{figure}

The final light curves of the target stars are  shown in
Fig. \ref{fig:lightcurves}. After detrending the light curves, we
calculated their Fourier transforms (FTs) to search for periodic signals.
The FT of the resulting light curves are shown in Fig. \ref{fig:FT}. 
For pre-whitening, we employed a nonlinear least square (NLLS) method, by simultaneously fitting each pulsation frequency above the 0.1\% false-alarm probability (FAP),
calculated randomizing the light curve 1000 times and measuring the highest peaks in their Fourier transforms. 
This iterative process has been done starting with the highest amplitude peak, 
until there is no peak that appears above $0.1 \%$ FAP significance threshold. 
All prewhitened frequencies for both targets are given in Table \ref{table:333432673} and \ref{table:095332541} including frequencies (periods) and amplitudes with their corresponding errors and the S/N ratio. 

In the case of TIC\,333432673, we detected 6 peaks, which are located between 2000 $\mu$Hz and 3000 $\mu$Hz. Two frequencies at 2015.393 and 2194.806 $\mu$Hz show significant residuals. For these unresolved peaks, we did not produce NLLS fit to extract from the light curves as they can be due to either photon-count noise caused by contamination of the background light in the aperture or amplitude/frequency/phase variations over the length of the data.
The frequencies of 2013.329 and 2015.393 $\mu$Hz can be considered as a rotationally split dipole mode.
If we assume that a central azimuthal component ($m = 0$) is missing, then the rotation period of TIC\,333432673 would be 5.6 d. If one of the side components ($m = +1$ or $-1$) is missing, then the rotation period of TIC\,333432673 would be 2.8 d.

In the case of TIC\,095332541, we identified 6 frequencies in the 120~s short-cadence (SC) data, while we detected 7 frequencies in 20~s ultra-short-cadence (USC) data. 
The frequencies that located in a similar region as TIC\,333432673, between 2200 and 2900 $\mu$Hz as can be seen in Fig. \ref{fig:FT}. 
In table \ref{table:095332541}, we present the frequency solution that is derived from the USC observations. All frequencies except 2403.783 $\mu$Hz (f$_{\rm 4}$) were found in the FT of SC data as well. 
The amplitudes of frequencies that were extracted from USC observations are about 2\% higher than the amplitudes of the SC observations.
This effect can be seen in the lower panel of Fig. \ref{fig:FT}, where we depicted FT of USC data (orange lines) and the FT of SC data (black lines). 
We found a doublet for TIC\,095332541 as well at 2401.602 and 2403.783 $\mu$Hz.
Again, if we assume that this pattern is due to rotational multiplets, then the rotation period of TIC\,095332541 would be either 2.65 d (in case of either $m = +1$ or $-1$ is missing) or 5.3 d (in case of a central azimuthal component $m = 0$ is missing).

The periodicities detected in these two PG~1159 stars are concentrated in a short region from 350\,s to 500\,s, which agree with the period spectrum typically exhibited by pulsating PG~1159 or GW Vir stars \citep[e.g.,][]{1979wdvd.coll..377M,1991ApJ...378..326W,2008A&A...477..627C, 2021A&A...645A.117C}. 
Due to the absence of evidence of any nebulae, these two GW Vir stars are classified as DOV stars. 

\begin{table*}
  \caption{The two new GW Vir stars reported from {\it
      TESS} observations,  including the name of
    the targets, {\it Gaia} magnitude, observed sectors,  date, \textit{CROWDSAP} estimate,  and length of the runs 
    (columns 1, 2, 3, 4, 5 and 6, respectively). From the Fourier Transform of the original and shuffled data, three different set of parameters: resolution, average
    noise level of amplitude spectra, and  detection threshold which
    we define as the amplitude at $0.1\%$ false alarm probability, FAP, are presented
    in columns 7, 8 and 9, respectively. The \textit{CROWDSAP} keyword shows the ratio of the target flux to the total flux in the optimal \textit{TESS} light curve aperture,
    taking into account that the TESS pixels are 21x21 arcsecs.}
  \begin{tabular}{ccccccccc}
\hline
Name &  $G_{\rm mag}$ & Obs.   & Start Time        &  \tt{CROWDSAP} & Length & Resolution & Average Noise & $0.1\%$\,FAP \\
     &                & Sector & (BJD-2\,457\,000) &   & [d]    & $\mu$Hz    &  Level [ppt]  &    [ppt]  \\
\hline
TIC\,333432673  & 15.66 & 33     & 2201.7372 & 0.14 & 25.83  &  0.45  & 0.95 & 3.67 \\
TIC\,095332541  & 15.79 & 33     & 2201.7373 & 0.78 & 25.83  &  0.45  & 0.71 & 3.06 \\
\hline
\label{DOVlist}
\end{tabular}
\end{table*}

\begin{table}
\centering
\caption{Independent frequencies, periods, and 
amplitudes, their uncertainties, and the 
signal-to-noise ratio in the  data of TIC\,333432673.
Errors are given in parenthesis to 2 significant digits.}
\begin{tabular}{ccccr}
\hline
\noalign{\smallskip}
Peak &$\nu$      & $\Pi$  &  $A$   & S/N \\
& ($\mu$Hz) &  (s)   & (ppt)  &   \\
\noalign{\smallskip}
\hline
\noalign{\smallskip}
f$_{\rm 1}$ &  2013.329 (44)	& 496.689 (10)	& 4.155   (75)	& 4.38    \\
f$_{\rm 2}$ &  2015.393 (23)	& 496.181 (05)	& 8.718   (77)	& 9.19    \\
f$_{\rm 3}$ &  2194.806 (15)	& 455.621 (03)	& 15.819  (80)	& 16.68   \\
f$_{\rm 4}$ &  2534.342 (47)	& 394.579 (07)	& 3.829   (75)	& 4.03    \\
f$_{\rm 5}$ &  2667.136 (29)	& 374.933 (04)	& 6.143   (75)	& 6.48    \\
f$_{\rm 6}$ &  2983.709 (33)	& 335.153 (03)	& 5.478   (75)	& 5.77    \\
\noalign{\smallskip}
\hline
\end{tabular}
\label{table:333432673}
\end{table}

\begin{table}
\centering
\caption{Independent frequencies, periods, and 
amplitudes, their uncertainties, and the 
signal-to-noise ratio in the  data of TIC\,095332541.
Errors are given in parenthesis to 2 significant digits.}
\begin{tabular}{ccccr}
\hline
\noalign{\smallskip}
Peak &$\nu$      & $\Pi$  &  $A$   & S/N \\
& ($\mu$Hz) &  (s)   & (ppt)  &   \\
\noalign{\smallskip}
\hline
\noalign{\smallskip}
f$_{\rm 1}$ &	2235.669 (27)	& 447.293 (55) &  4.975	 (57)  & 7.10  \\ 
f$_{\rm 2}$ &	2264.722 (37)	& 441.555 (72) &  3.675	 (57)  & 5.25  \\
f$_{\rm 3}$ &	2401.602 (21)	& 416.389 (37) &  6.447	 (57)  & 9.21  \\
f$_{\rm 4}$ &	2403.783 (44)	& 416.011 (77) &  3.105	 (57)  & 4.43  \\
f$_{\rm 5}$ &	2437.256 (38)	& 410.298 (64) &  3.578	 (57)  & 5.11  \\
f$_{\rm 6}$ &	2472.863 (09)	& 404.390 (15) &  15.119 (57)  & 21.59  \\
f$_{\rm 7}$ &	2836.338 (11)	& 352.567 (13) &  12.856 (57)  & 18.36  \\
\noalign{\smallskip}
\hline
\end{tabular}
\label{table:095332541}
\end{table}

\section{Asteroseismology}
\label{ast_analysis}

\begin{figure} 
\includegraphics[clip,width=1.\linewidth]{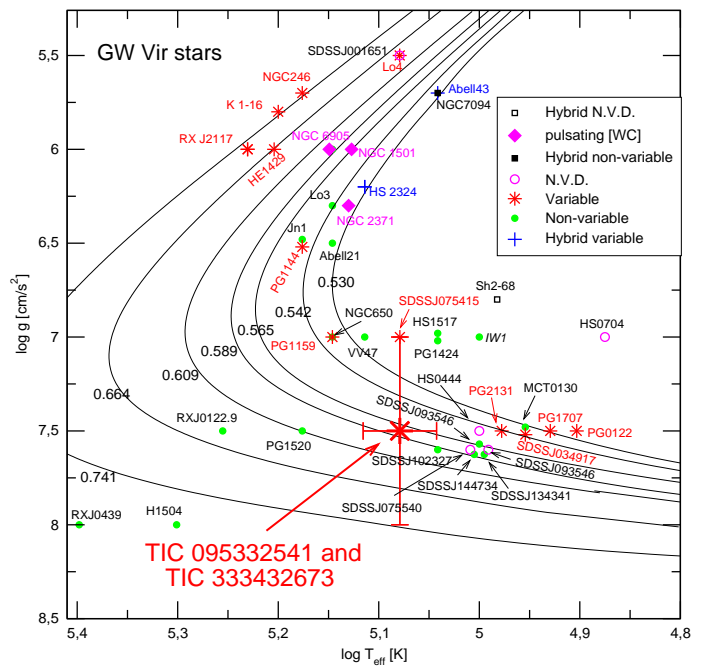}
\caption{The already known variable and
  non-variable PG~1159 stars and variable [WCE] stars 
  in the $\log T_{\rm eff}-\log g$ plane.
  Thin solid  curves  show  the post-born again evolutionary tracks
  from \cite{2006A&A...454..845M} for different stellar masses.  
  "N.V.D." stands for PG~1159 stars  with no variability data. "Hybrid" refers
  to PG~1159 stars exhibiting H in their atmospheres. The location of the 
  two new GW Vir stars TIC~333432673 and TIC~095332541 
  is emphasized with  a large red star symbol with error bars. 
  Both stars share the same 
  spectroscopic surface parameters, $T_{\rm eff}= 120\,000\pm 10\,000$ K and 
  $\log g= 7.5\pm0.5$.}
\label{fig:logTefflogg} 
\end{figure} 

\subsection{Spectroscopic mass}
\label{spectroscopic-mass}

For the asteroseismological analysis of this work, we employ the set of  
state-of-the-art evolutionary models of PG~1159 stars of  \cite{2005A&A...435..631A}
and \cite{2006A&A...454..845M,2007A&A...470..675M,2007MNRAS.380..763M}.  
In those works, a set of post-AGB evolutionary sequences  computed with the {\tt
  LPCODE} evolutionary code \citep{2005A&A...435..631A} were followed
through  the  very  late   thermal  pulse  (VLTP)  and  the  resulting
born-again  episode that  give rise  to the  H-deficient, He-,  C- and
O-rich composition characteristic of  PG~1159 stars.  The masses of the
resulting  remnant models are  $0.530$, $0.542$, $0.565$, $0.589$, $0.609$,
$0.664$, and  $0.741 \ M_{\sun}$. In Fig.~\ref{fig:logTefflogg} we show the
evolutionary tracks of PG~1159 stars in the $\log
T_{\rm eff}$ vs. $\log g$ plane. In Table \ref{tabla-grilla-modelos} we show 
the values of the stellar mass, the mass of the envelope, and the surface chemical 
abundances by mass of ${^4}$He, $^{12}$C, and $^{16}$O  for the evolutionary sequences 
employed  in  this  study.   $M_{\rm env}$ is defined as the mass external 
to the location of the O/C/He chemical transition region. Since this chemical 
transition has a finite width (see Fig. \ref{asteroseismological-model}), 
we consider the location of the interface as the point at which 
the He abundance has reduced to half of its surface abundance.
We note that, in this work, element diffusion was neglected. This is motivated by theoretical
expectations that residual weak winds and radiative acceleration retard
the gravitational settling of He until the surface gravity reaches a value
between $\log g = 7.5$ and $8$ \citep[see][]{2000A&A...359.1042U}. For a $0.529 \ M_{\sun}$ PG~1159 model, this occurs by $T_{\rm eff} = 65\,000$~K. At higher $T_{\rm eff}$, the abundances of CNO elements remain nearly unchanged.

We emphasise the need of evolutionary models derived from the full
computation of the late thermal pulse to perform seismological studies of PG~1159 stars. Such models provide realistic predictions of the internal chemical stratification and a correct assessment of the evolutionary time scales during the PG 1159 regime, where He shell burning is the main source of energy.
The set of PG 1159  sequences we use in this work reproduces the
spread in surface chemical composition observed in PG 1159 stars, the short
born-again time of post born again objects,  and the location of the
GW Vir instability strip, see \cite{2006A&A...458..259C,2010A&ARv..18..471A}.
During the very late thermal pulse, an outward-growing convection zone driven by
the He-burning shell develops and reaches the H-rich envelope. As a result,
most  of the H content of the remnant is violently burned in the He-flash convection zone. After this short-lived evolutionary stage, during which the remnant
returns from the hot white dwarf stage to the red giant state, evolution
proceeds to the domain of PG 1159 stars with a H-deficient surface composition
rich in He, C and O. The interplay between  mixing and burning during the
late thermal pulse creates large quantities  of  $^{13}$C and $^{14}$N. The role of such isotopes is by no means negligible since during the PG 1159 regime, $\alpha$  captures by  $^{13}$C affect the shape of the O profile at the base of the He, C, O envelope (see Fig. \ref{asteroseismological-model}).

\begin{table}
\centering
\caption{Stellar mass (in solar units), surface gravity, envelope mass, and the 
surface chemical abundances by mass for the evolutionary 
sequences considered in this work. The values correspond to models with 
$\log(T_{\rm eff})= 5.1$ (K) at the stages before the evolutionary knee,
that is, the maximum effective temperature possible 
(see Fig. \ref{fig:logTefflogg}).}
\begin{tabular}{lccccc}
\hline
\hline
 $M_{\star}/M_{\odot}$ & $\log g$ (cgs) &$\log(M_{\rm env}/M_{\star})$ & $X_{^4{\rm He}}$ & $X_{^{12} {\rm C}}$ & $X_{^{16}{\rm O}}$ \\
\hline
 0.530 &  6.120 & $-1.18$ & 0.33  &  0.39  & 0.17  \\
 0.542 &  5.944 & $-1.22$ & 0.28  &  0.41  & 0.21  \\ 
 0.565 &  5.853 & $-1.34$ & 0.39  &  0.27  & 0.22  \\
 0.589 &  5.780 & $-1.40$ & 0.31  &  0.38  & 0.23  \\
 0.609 &  5.677 & $-1.51$ & 0.50  &  0.35  & 0.10  \\
 0.664 &  5.541 & $-1.86$ & 0.47  &  0.33  & 0.13  \\
 0.741 &  5.417 & $-2.00$ & 0.48  &  0.34  & 0.14  \\
\hline
\end{tabular}
\label{tabla-grilla-modelos}
\end{table}  

On   the    basis   of   the   
evolutionary    tracks   and the values of the spectroscopic  
surface gravity and effective temperature of TIC~333432673 and TIC~095332541, 
we derive by interpolation a value of the spectroscopic  mass.  
We obtain a stellar mass of  $M_{\star}= 0.58^{+0.16}_{-0.08}
\ M_{\sun}$ for both GW Vir stars. The large uncertainties of the spectroscopic 
mass come mainly from the uncertainties in the surface gravity. 

\subsection{Mean period spacing of TIC~333432673}
\label{mean_period_spacing-TIC333}

In the asymptotic limit of stellar pulsations, i.e.,  for large radial orders 
($k \gg \ell$), $g$ modes of consecutive radial 
order in WDs and pre-WDs are approximately uniformly
spaced in period \citep{1990ApJS...72..335T}. The asymptotic period spacing
is given by $\Delta \Pi_{\ell}^{\rm a} = {\Pi}_{0}/{\sqrt{\ell(\ell+1)}}$,
$\Pi_{0}$ being a constant defined as
$\Pi_{0}=  2 \pi^2 \left[\int_{r_1}^{r_2}
\frac{N}{r} dr \right]^{-1}$, where $N$ is the Brunt-V\"ais\"al\"a frequency
\citep[see, e.g.,][for its definition]{1989nos..book.....U}. 
This asymptotic formula constitutes a very precise description of the
pulsational properties of chemically homogeneous stellar models. 
However, the  $g$-mode period spacings in chemically
stratified PG~1159  stars show  appreciable departures from uniformity
caused by the mechanical resonance  called "mode trapping" 
\citep{1994ApJ...427..415K}. The observed {\it average}
period spacing of GW Vir stars primarily depends on the stellar mass and 
the effective temperature \citep{1994ApJ...427..415K}. 
It allows us to estimate $M_{\star}$ by fixing $T_{\rm eff}$. 
Specifically, a way to derive an estimate of the stellar mass is by comparing the
observed average period spacing ($\Delta \Pi$) of the target star with the
asymptotic period spacing ($\Delta \Pi_{\ell}^{\rm a}$) computed 
at the effective temperature of the star \citep[see the pioneer works of ][]{1987fbs..conf..297K,1988IAUS..123..329K}. 
Since GW Vir stars generally do not have all of their 
pulsation modes in the asymptotic regime, there is usually no perfect 
agreement between $\Delta \Pi$ and $\Delta \Pi_{\ell}^{\rm a}$.    
A variation of this approach is to compare $\Delta \Pi$ with
the {\it average} of the computed period spacings ($\overline{\Delta
  \Pi_{k}}$), instead the asymptotic period spacing. 
  
\begin{figure} 
\includegraphics[clip,width=1.0\columnwidth]{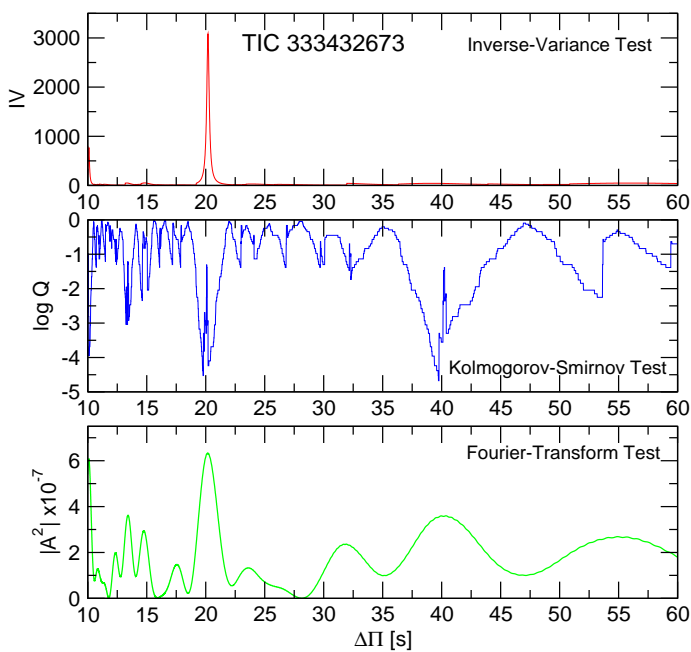}
\caption{I-V  (upper panel), K-S  (middle panel), and F-T  
significance  tests  to  search  for  a constant  period
  spacing  in TIC~333432673. The tests are applied to the set of 6
  pulsation periods of Table \ref{table:333432673}. A 
  strong signal of a constant period spacing at  $\sim 20$~s is evident.
  See text for details.}
\label{test-TIC333432673} 
\end{figure}

\begin{figure} 
\includegraphics[clip,width=1.0\columnwidth]{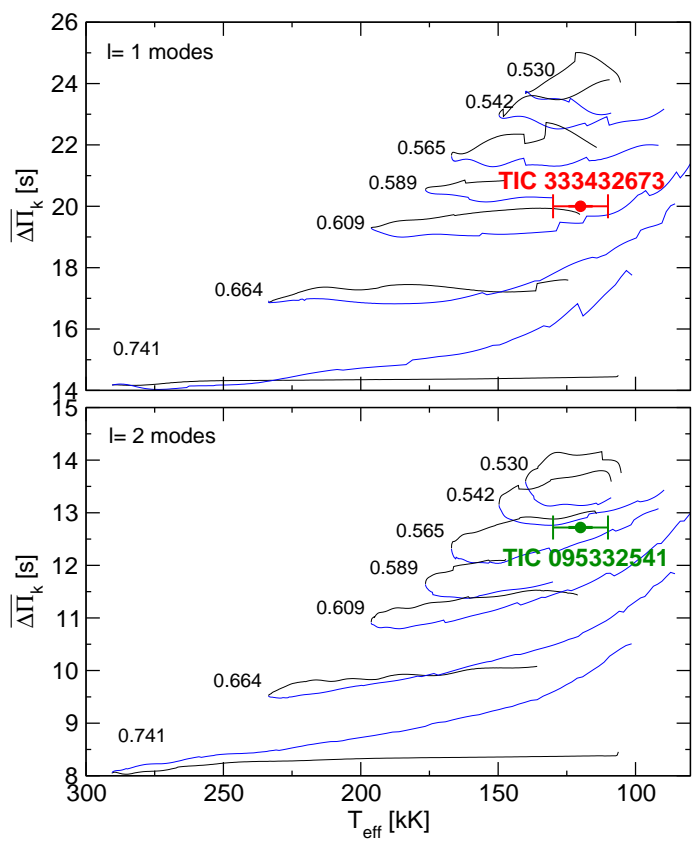}
\caption{Upper panel: dipole ($\ell= 1$) average of the computed
  period spacings, $\overline{\Delta \Pi_k}$, assessed  in  a  range
  of  periods  that includes  the  periods  observed  in the GW Vir star TIC~095332541 
  and TIC~333432673, shown as black (blue) curves corresponding to stages before
  (after) the maximum $T_{\rm eff}$ for different stellar masses. The
  location of TIC~333432673 when we  use the spectroscopic effective temperature,  
  $T_{\rm eff}= 120\,000 \pm 10\,000$ K, and the dipole period spacing 
  $\Delta \Pi_{\ell= 1}= 20.19$~s, is highlighted
  with a red circle in the upper panel.  Lower panel: same as in upper panel,  but for the 
  average of the computed period spacings with $\ell= 2$. The GW Vir star TIC~09533254 
  is drawn in this plot considering a quadrupole period spacing  of $\sim 13$ s (see Sect. \ref{mean_period_spacing-TIC095}).}
\label{fig:period_spacing_teff} 
\end{figure}

We  searched for a  constant  period  spacing  in  the  data of
the new GW Vir stars  using the  Kolmogorov-Smirnov
\citep[K-S;][]{1988IAUS..123..329K}, the inverse  variance
\citep[I-V;][]{1994MNRAS.270..222O}, and the Fourier Transform
\citep[F-T;][]{1997MNRAS.286..303H} significance tests.  In 
Fig. \ref{test-TIC333432673} we show  the results of applying these
tests to the set of 6 periods of TIC~333432673 (Table \ref{table:333432673}). 
A very strong signature of a period spacing of $\sim 20$ s is evident 
according to the three tests. There is also an indication of a 
possible period spacing of $\sim 40$ s according to the K-S 
and F-T tests, although it is completely absent in the I-V test. 
In Fig. \ref{fig:period_spacing_teff} we show
the dipole ($\ell= 1$, upper panel) and quadrupole ($\ell= 2$, lower panel) 
average of the computed period spacings, $\overline{\Delta \Pi_k}$, 
assessed  in  a  range  of  periods  that includes  the  periods  
observed  in TIC~095332541 and TIC~333432673 ($300-500$ s), shown  as 
curves for different stellar masses. The $g$-mode  adiabatic pulsation  
periods employed to assess $\overline{\Delta \Pi_k}$ 
were computed with  the  {\tt LP-PUL} pulsation code \citep[][]{2006A&A...454..863C}. 
We can safely discard a period spacing of $\sim 40$\,s  
due to the fact that if such a long period spacing were real, TIC~333432673 
should have an extremely low mass, irreconcilable with the 
spectroscopic mass. Thus,  we can assume that the mean period spacing of 
$20.19$ s is robust and reliable for this star. It  can be 
associated with a  sequence of $\ell= 1$ 
modes\footnote{If they were $\ell= 2$, then the stellar mass 
of TIC~333432673 would be extremely low (see lower panel of Fig. \ref{fig:period_spacing_teff}), and thus, 
incompatible with the spectroscopic mass.}. An $\ell= 2$ period spacing, if present, 
should have a value $\Delta \Pi_{\ell= 2} \sim \Delta \Pi_{\ell= 1}/\sqrt{3}\sim 11.7$ s. 
Given the absence of a period spacing of this value in the three statistical tests, we 
conclude that there is no period spacing corresponding to modes with $\ell= 2$ in the observed
pulsational spectrum of TIC~333432673.
We compared the period spacing of $\sim 20.19$ s with the 
$\overline{\Delta \Pi_{k}}$ in terms of $T_{\rm eff}$ for all the
masses considered in the upper panel of Fig.~\ref{fig:period_spacing_teff}. 
The resulting mass value is $M_{\star} \sim 0.61\,M_{\odot}$ or $M_{\star}\sim 0.60\, M_{\odot}$ if the star is before or after the
evolutionary knee, respectively. These values are in good
agreement with the spectroscopic mass value ($M_{\star}= 0.58^{+0.16}_{-0.08}
\ M_{\sun}$). 

\subsection{Period-to-period fits for TIC~333432673}
\label{period-to-period-fits-TIC333}

A powerful asteroseismological tool to disentangle the internal structure
of GW Vir stars is to seek theoretical models that best match the
individual pulsation  periods  of  the target stars. To measure the
goodness of the match between the theoretical pulsation periods
($\Pi_{\ell,k}$) and the observed individual periods ($\Pi_i^{\rm
  o}$), we assess the merit function  $\chi^2(M_{\star}, T_{\rm eff})= \frac{1}{N} \sum_{i= 1}^{N} {\rm min}[(\Pi_{\ell,k}-\Pi_i^{\rm o})^2]$ \citep[see, for instance,][]{2021A&A...645A.117C}. Here, $N$ is the number of observed periods.
  In order to find the stellar model that best replicates the observed periods exhibited by
each target star --- the ``asteroseismological'' model --- we
evaluate the  function  $\chi^2$ for stellar  masses between  
$0.530$ and $0.741 M_{\odot}$ and effective temperatures in the range 
$80\,000-300\,000$ K, with $\Delta T_{\rm eff} \sim 30$\,K.  For each 
target star, the PG~1159 model that shows the lowest value of $\chi^2$ is adopted as the best-fit asteroseismological model. 

\begin{figure} 
\includegraphics[clip,width= 1.0\columnwidth]{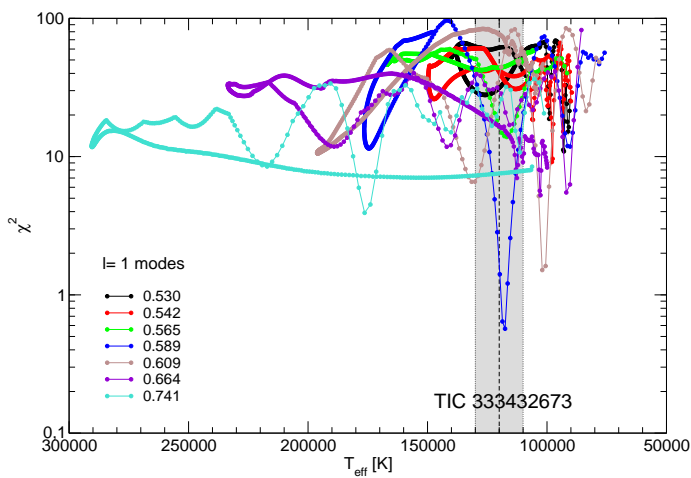}
\caption{The quality  function  of the  period  fits in terms of  the
  effective temperature  for the PG~1159 sequences with different
  stellar masses  (in solar units) corresponding to the case 
  in which all the observed periods are assumed to be associated 
  to dipole ($\ell= 1$) modes. We note  the presence of a
  strong minimum corresponding  to $M_{\star}= 0.589 M_{\odot}$ and 
  $T_{\rm eff}= 117\,560$ K.   The vertical dashed line
  is the spectroscopic  $T_{\rm eff}$ of  TIC~333432673  (120\,000~K)  and  the
  gray zone depicts its uncertainties  ($\pm 10\,000$~K).}
\label{fig:chi2-TIC333432673} 
\end{figure} 

\begin{figure} 
\includegraphics[clip,width= 1.0\columnwidth]{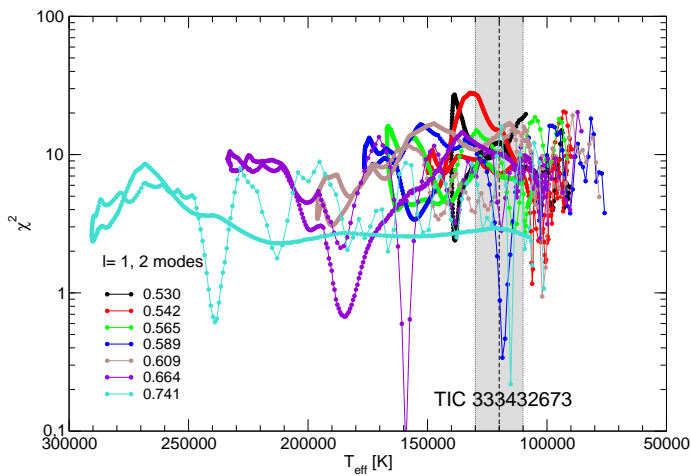}
\caption{The quality  function  of the  period  fits in terms of  the
  effective temperature  for the PG~1159 sequences with different
  stellar masses  (in solar units) corresponding to the case in which 
  the observed periods are assumed to be associated to dipole ($\ell = 1$) and 
  quadrupole ($\ell= 2$) modes. We note  the presence of two
  strong minimum compatible with the effective temperature of the star, 
  corresponding  to $M_{\star}= 0.589 M_{\odot}$ and $T_{\rm eff}= 118\,655$ K,
 and  $M_{\star}= 0.741 M_{\odot}$ and $T_{\rm eff}= 115\,203$ K.}
\label{fig:chi2-TIC333432673-l1l2} 
\end{figure}

\begin{table}
\centering
\caption{Observed and theoretical periods of the asteroseismological
  model for TIC~333432673  [$M_{\star}= 0.589 M_{\odot}$, $T_{\rm eff}=
    117\,560$ K, $\log(L_{\star}/L_{\odot})= 1.896$].  Periods are in seconds and rates of period change (theoretical) are in units of $10^{-12}$ s/s.  $\delta \Pi_i= \Pi^{\rm O}_i-\Pi_k$ represents the period differences, $\ell$ the harmonic degree, and $k$
    the radial order.}
\begin{tabular}{cc|ccccc}
\hline
\noalign{\smallskip}
$\Pi_i^{\rm O}$ & $\ell^{\rm O}$ & $\Pi_k$ & $\ell$ & $k$ &  $\delta \Pi_k$ & $\dot{\Pi}_k$ \\
(s) & & (s) & & &  (s) & ($10^{-12}$ s/s)  \\
\noalign{\smallskip }
\hline 
\noalign{\smallskip}      
335.153 & 1 & 335.306 & 1 & 14 & $-0.153$ & 6.006  \\
374.933 & 1 & 374.215 & 1 & 16 & $ 0.718$ & 6.763  \\
394.579 & 1 & 396.035 & 1 & 17 & $-1.456$ & 5.165  \\
455.621 & 1 & 455.941 & 1 & 20 & $-0.320$ & 5.632  \\
496.181 & 1 & 496.467 & 1 & 22 & $-0.286$ & 8.181  \\
\noalign{\smallskip}
\hline
\end{tabular}
\label{table:TIC333432673-asteroseismic-model}
\end{table}

The period-to-period fits for TIC~333432673 lead us to an excellent seismological solution for this star. The quality function versus the effective temperature for the different stellar masses corresponding to the case in which we assume that all the observed periods correspond to $\ell= 1$ modes is shown in  Fig. \ref{fig:chi2-TIC333432673}. We note that, among the periods 496.181 s and 496.689 s\footnote{These two very close periods could be part of a rotational frequency triplet; see at the end of Sect. \ref{photometric_observations}, 
we have retained only the largest amplitude one, that is, 496.181 s. So,
we employ a set of 5 observed periods in our period fits (first column in Table \ref{table:TIC333432673-asteroseismic-model}).
It is noteworthy the existence of a very clear minima of the quality 
function ($\chi^2= 0.568$ s$^2$) corresponding to an asteroseismological model characterized by
an effective temperature of $T_{\rm eff}= 117\,560$ K, very close to the spectroscopic effective 
temperature of TIC~333432673 and well within its uncertainties ($T_{\rm eff}= 120\,000 \pm 10\,000$ K).}
The stellar mass of this model is $M_{\star}= 0.589\,M_{\sun}$, in 
perfect agreement with the mass derived with the period spacing, 
$M_{\star}= 0.60-0.61\,M_{\sun}$, and the spectroscopic mass value, 
$M_{\star}= 0.58\,M_{\sun}$. 

We have also performed period fits assuming the situation 
in which the 5 observed periods are a mix of $\ell= 1$ and $\ell= 2$ periods, i.e.,
we compare the observed periods with dipole and quadrupole theoretical periods. In this case, 
which is illustrated in Fig. \ref{fig:chi2-TIC333432673-l1l2},  we find two possible solutions
compatible with the spectroscopic effective temperature, one of them characterized by 
$T_{\rm eff}= 118\,655$ K, $M_{\star}= 0.589\,M_{\sun}$, and $\chi^2= 0.340$ s$^2$
and the other one characterized by $T_{\rm eff}= 115\,203$ K, 
$M_{\star}= 0.741\,M_{\sun}$ and $\chi^2= 0.219$ s$^2$. In addition, a very strong minimum of 
$\chi^2$ is found for a model with $T_{\rm eff}= 159\,100$ K, 
$M_{\star}= 0.664\,M_{\sun}$ and $\chi^2= 0.070$ s$^2$, 
but we have to discard this model as a possible solution since it is too hot 
and incompatible with the range of possible effective temperatures 
given by spectroscopy.

In summary, we face the problem of choosing an asteroseismological model among three possible solutions. In Section \ref{mean_period_spacing-TIC333} we found that 
the periods of TIC~333432673 make up an equispaced pattern with a period separation of $20.19$~s 
corresponding to dipole modes. This finding constitutes a strong constraint on the assignment 
of the $\ell$ value of the modes, that is, it indicates that the identification of all the
observed periods with modes $\ell= 1$ is robust. In \cite{2021A&A...645A.117C} we have employed this constraint to assign the $\ell$ value of the periods of a set of 
GW Vir stars observed with {\it TESS}. Using that constraint for TIC~333432673, we safely retain the solution characterized by $T_{\rm eff}= 117\,560$ K and $M_{\star}= 0.589\,M_{\sun}$, in which all modes are dipole modes, as the asteroseismological model for this star. Note that this solution is very close to one of the possible solutions obtained when we assume a possible mixture of $\ell= 1$ and $\ell = 2$ modes in the observed spectrum of TIC~333432673 ($T_{\rm eff}= 118\,655$ K and $M_{\star}= 0.589\,M_{\sun}$).

In Table~\ref{table:TIC333432673-asteroseismic-model} 
we show a  detailed comparison of the observed periods of TIC~333432673 and 
the theoretical periods  of  the  asteroseismological  model. According to our
asteroseismological model, all the  periods exhibited by the star
correspond to $\ell= 1$ modes with high radial order $k$. We compute the
average   of   the   absolute   period    differences,
$\overline{\delta \Pi_i}= \left(\sum_{i= 1}^n |\delta \Pi_i|
\right)/n$, where $\delta \Pi_i= (\Pi_{\ell,k}-\Pi_i^{\rm o})$ and $n= 5$,  
and the root-mean-square residual, $\sigma= \sqrt{(\sum_{i= 1}^n
  |\delta \Pi_i|^2)/n}= \sqrt{\chi^2}$.  We compute also the Bayes
Information Criterion \citep[BIC;][]{2000MNRAS.311..636K}, 
${\rm BIC}= n_{\rm p} \left(\frac{\log n}{n} \right) + \log \sigma^2$, 
where $n_{\rm p}$ is the number of free parameters of the
models, and $n$ is the number of fitted periods. The smaller the
value of BIC, the better the quality of the fit. In our case, $n_{\rm
  p}= 2$ (stellar mass and effective temperature), and  $n= 5$.  We obtain $\overline{\delta  \Pi_i}= 0.59$~s, $\sigma= 0.75$~s, and ${\rm BIC}= 0.03$, which means   
  that our period fit is excellent, although  admittedly, the differences 
  between theoretical and observed periods is larger than 
  the uncertainties in the measured periods. 

We also include in Table~\ref{table:TIC333432673-asteroseismic-model} the
rates of period change ($\dot{\Pi}\equiv d\Pi/dt$) predicted for each
$g$ mode of TIC~333432673. Note that all of them are positive
($\dot{\Pi}>0$), implying that the periods are lengthening  over
time. The rate of  change of periods in WDs and pre-WDs is related
to $\dot{T}$ ($T$ being the temperature at the region of the period
formation) and $\dot{R_{\star}}$ ($R_{\star}$ being the stellar
radius) through the order-of-magnitude expression $(\dot{\Pi}/\Pi) \approx -a\ (\dot{T}/T) + b\ (\dot{R_{\star}}/R_{\star})$
\citep{1983Natur.303..781W}. According  to our asteroseismological
model, the star is cooling at almost constant radius after reaching 
its maximum temperature (evolutionary knee), i.e., TIC~333432673 is in its 
cooling stage. As a consequence, $\dot{T}<0$ and $\dot{R_{\star}} \sim 0$, and then, $\dot{\Pi} > 0$. Continuous monitoring of this star could in the 
future make it possible to measure rates of period change for comparison with our theoretical predictions, if the pulsations are shown to be otherwise coherent over the span of observations.

\begin{table}
\centering
\caption{The main characteristics of the new GW Vir star TIC~333432673. The second column  
  corresponds to spectroscopic results, whereas the third  column present results
  from the asteroseismological model.} 
\begin{tabular}{l|cc}
\hline
\hline
Quantity & Spectroscopy &  Asteroseismology \\
         & Astrometry   &         \\ 
\hline
$T_{\rm eff}$ [kK]                          & $120 \pm 10$                        & $118\pm12$ \\
$M_{\star}$ [$M_{\odot}$]                   & $0.58^{+0.16}_{-0.08}$              & $0.589\pm 0.020$ \\ 
$\log g$ [cm/s$^2$]                         & $7.5\pm0.5$                         & $7.55_{-0.55}^{+0.52}$  \\ 
$\log (L_{\star}/L_{\odot})$                & $\hdots$                            & $1.90^{+0.25}_{-0.34}$ \\  
$\log(R_{\star}/R_{\odot})$                 & $\hdots$                            & $-1.67^{+0.22}_{-0.25}$ \\  
$(X_{\rm He},X_{\rm C}, X_{\rm O})_{\rm s}$ & 0.75, 0.25, $\cdots$                & 0.30, 0.38, 0.23\\
$d$  [pc]                                   & $389\pm 5^a$                           & $459^{+188}_{-156}$   \\ 
$\pi$ [mas]                                 & $2.57^{+0.07,a}_{-0.04}$    & $2.18_{-1.85}^{+1.12}$\\ 
\hline
\hline
\end{tabular}
\label{table:modelos-sismo-TIC333432673}

{\footnotesize  References: (a) {\it Gaia DR3}.}
\end{table}


In Table~\ref{table:modelos-sismo-TIC333432673}, we list the 
spectroscopic and astrometric parameters of TIC~333432673 and 
the main characteristics of the asteroseismological model found 
in this work. The seismological stellar mass is in excellent agreement 
with the value derived from the period 
spacing ($0.60-061\, M_{\odot}$) and the spectroscopic mass. 
The average of the dipole ($\ell= 1$) period spacings of
our  asteroseismological model is $\overline{\Delta \Pi}= 20.469$ s,
in excellent agreement with the  $\ell= 1$ mean period spacing derived
for TIC~333432673, $\Delta \Pi= 20.19$~s.  

\begin{figure} 
\includegraphics[clip,width=1.0\columnwidth]{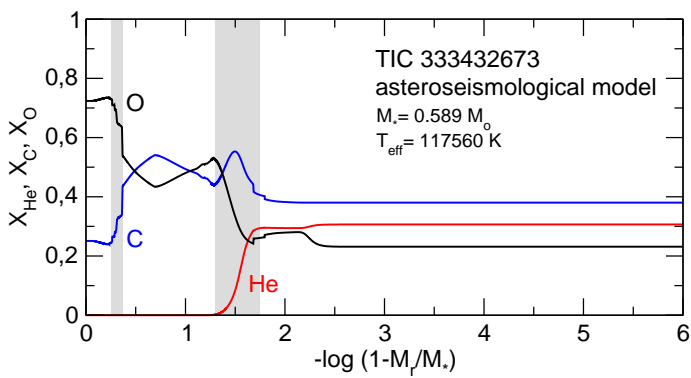}
\caption{The  internal  chemical  profile  of  the asteroseismological model of TIC~333432673  
($M_{\star}= 0.589 M_{\sun}$, $T_{\rm eff}= 117\,560$ K)  in  terms  of  the  
outer  fractional  mass.  The locations of the O/C and O/C/He chemical interfaces
are indicated with gray regions.}
\label{asteroseismological-model} 
\end{figure}

In Fig. \ref{asteroseismological-model} we show the fractional abundances of the main chemical species, $^4$He, $^{12}$C, and $^{16}$O, corresponding to our asteroseismological model with $M_{\star}= 0.589 M_{\odot}$  and $T_{\rm eff}= 117\,560$ K. The chemical transition regions of O/C and O/C/He are clearly visible. The location, thickness, and steepness of these chemical interfaces define the mode-trapping properties of the models.  \citep[see, e.g.,][for details]{2005A&A...439L..31C,2006A&A...454..863C}.

\subsection{The asteroseismological distance of TIC~333432673}
\label{asteroseismological_distance-TIC333}

We can assess the asteroseismological distance  on the basis of the 
luminosity of the asteroseismological model [$\log(L_{\star}/L_{\odot})= 1.90^{+0.25}_{-0.34}$]. 
Using a bolometric correction $BC= -7.05$, interpolated from the values 
corresponding to PG~1159$-$035 ($T_{\rm eff}= 140\,000$~K, $BC= -7.6$), 
RX~J2117+3142 ($T_{\rm eff}= 160\,000$ K, $BC= -7.95$), and PG~2131+066 ($T_{\rm eff}= 95\,000$ K, 
$BC= -6.0$), as given by \cite{1994ApJ...427..415K}, \cite{2007A&A...461.1095C}, and \cite{1995ApJ...450..350K}, respectively,
the visual absolute magnitude can be assessed as $M_{\rm V}= M_{\rm B}-BC$,  where $M_{\rm
  B}= M_{{\rm B},\odot} - 2.5\ \times \log(L_{\star}/L_{\odot})$. We employ
the solar bolometric magnitude $M_{\rm B \odot}= 4.74$
\citep{2000asqu.book.....C}. The seismological distance $d$  is
derived from the relation: $\log d= (1/5)\ [m_{\rm V} - M_{\rm V} +5 -
  A_{\rm V} (d)]$,  where we employ the 3D reddening map of \cite{2014A&A...561A..91L} \citep[see also][]{2017A&A...606A..65C,2018A&A...616A.132L}\footnote{\tt \url{https://stilism.obspm.fr/}} to infer $E(B-V)$ and then the interstellar  absorption $A_{\rm V}(d)$, which  
  is a nonlinear function of the  distance  and  also  depends
on  the  Galactic  latitude  ($b$) and longitude ($l$).  For  the  equatorial  coordinates
of  TIC~333432673  (Epoch  B2000.00, $\alpha= 6^{\rm h} 41^{\rm m} 15.^{\rm
  s}88,\ \delta= -13^{\circ} 41^{'} 31.^{''}31$) the corresponding
Galactic latitude is $b=-8.^{\circ}41280066$ and $l=224^{\circ}.06462259$.  We use the
apparent visual magnitude  $m_{\rm V}= 15.658$ ({\it TESS} catalog),  
and obtain iteratively the seismological distance, $d= 459^{+188}_{-156}$ pc, 
parallax  $\pi= 2.18^{+1.12}_{-1.85}$~mas, and extinction coefficient  $A_{\rm V}=
0.307^{+0.053}_{-0.121}$. The large uncertainty in the seismological distance comes
mainly from the large uncertainty in the  luminosity of the
asteroseismological model and in the reddening coefficient $E(B-V)$. 
A very important check for the validation of the
asteroseismological model of TIC~333432673 is the comparison of the
seismological  distance with the distance derived from astrometry. We
have available the estimates from {\it Gaia EDR3} \citep{2021A&A...650C...3G}, $d_{\rm G}= 389.0^{+5.6}_{-5.2}$ 
pc and $\pi_{\rm G}= 2.57^{+0.07}_{-0.04}$ mas.  They are in excellent
agreement with the asteroseismological derivations in view of the
uncertainties in both determinations. This adds confidence to the 
correctness of the asteroseismological model.

\subsection{Mean period spacing of TIC~095332541}
\label{mean_period_spacing-TIC095}

\begin{figure} 
\includegraphics[clip,width=1.0\columnwidth]{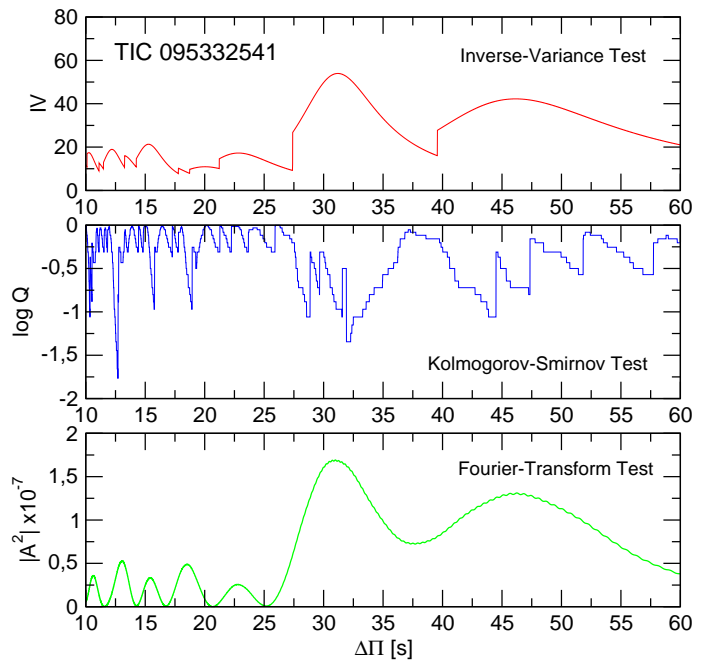}
\caption{I-V  (upper panel) and  K-S  (middle panel), and F-T (bottom panel)  
significance  tests  to  search  for  a constant  period
  spacing  in TIC~095332541, applied to the set of 7
  pulsation periods of Table \ref{table:095332541}. A 
  signal of a constant period spacing at  $\sim 31$~s is evident 
  in the three tests.  See text for details.}
\label{test-TIC095332541} 
\end{figure} 

In Fig. \ref{test-TIC095332541}  we show  the results of applying 
the significance tests to the set of 7 periods of TIC~095332541
(Table \ref{table:095332541}). The three tests point to the existence
of a probable constant period spacing of $\Delta \Pi\sim
31$~s. If we want to guess what the stellar mass 
corresponding to this period spacing would be by comparing it with the average of
the computed period spacings, we realize that we cannot draw the location of TIC~095332541 in the diagrams of Fig. \ref{fig:period_spacing_teff}. A period spacing as long as $\sim 31$ s for TIC~095332541 does not 
make physical sense, since it would be compatible with a star with unusually low mass (much below $0.4 M_{\odot}$), in strong conflict with the spectroscopic 
mass. Therefore, this possible period spacing has to be disregarded. 
For the same reason, we also rule out a possible period spacing of $\sim 45$ s
that can be seen in Fig. \ref{test-TIC095332541}. 
A third possible constant period separation is $\sim 13$ s,  
as suggested by the K-S test. Such a short period spacing could make any sense 
only if all the modes exhibited are quadrupole ($\ell= 2$) modes. 
In this case, the stellar mass would be $\sim 0.55 M_{\odot}$ 
(see lower panel of Fig. \ref{fig:period_spacing_teff}). 
However, since this possible period-spacing signature is only suggested 
by one of the significance tests, we cannot take it as true, and we are 
forced to discard it. We conclude that it is not possible to find a realistic 
period spacing for TIC~095332541 with this data set of only 7 detected periods. 

\begin{figure} 
\includegraphics[clip,width=1.0\columnwidth]{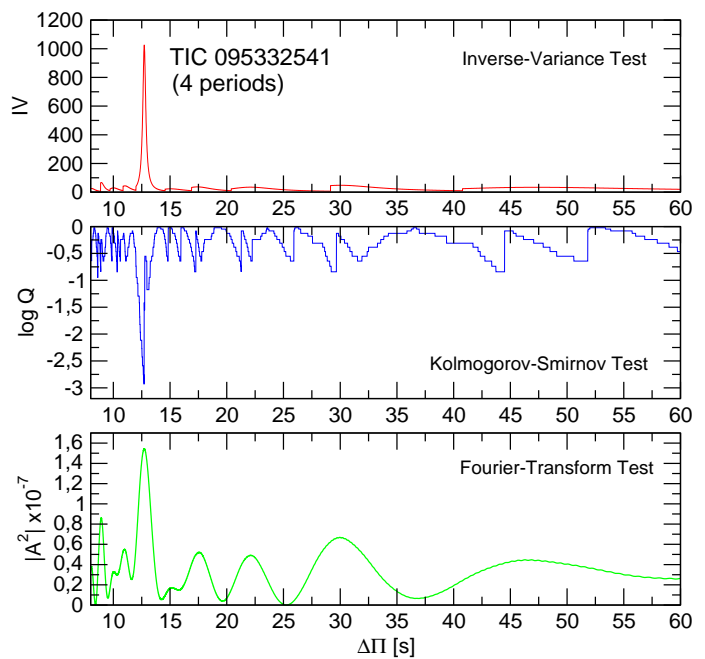}
\caption{Same as in Fig. \ref{test-TIC095332541}, but for the case in which we consider the subset of 4 periods 352.567 s, 404.390 s, 416.389 s, 
and 441.555 s. See text for details.}
\label{test-TIC095332541-4p} 
\end{figure}

We have sought an alternative interpretation of the period spectrum of TIC~095332541 
in which we consider subsets of periods when looking for patterns of period spacing. 
In particular, if we consider the subset of 4 periods 352.567 s, 404.390 s, 416.389 s, and 441.555 s, 
we find what is plotted in Fig. \ref{test-TIC095332541-4p}, where we show the statistical tests considering 
this reduced set of periods. The existence of a period spacing of around $\sim 13$ s is evident according the three significance tests. Averaging the results between the three tests, we find $\Delta \Pi= 12.71$ s. A period spacing of $\sim 13$ s can be due to
that the four periods considered are associated to $\ell= 2$ modes, which implies a 
stellar mass in the range $\sim 0.548 - 0.570 M_{\odot}$ (see lower panel of Fig. \ref{fig:period_spacing_teff}).  The remainder periods (410.298 s, and 447.293\footnote{The period at 416.011 s is very similar to the period at 416.389 s, and could be a component of a rotational multiplet or a single
mode with $\ell = 1$. We do not consider it in the computation of the period spacing nor in our period fits; see the next section.}) could be $\ell= 2$ modes affected by mode trapping, or, 
alternatively, $\ell= 1$ modes.  The absence of a 
dipole period spacing at $\sim 21$ sec (according to the asymptotic theory) and the simultaneous presence of a quadrupole period spacing at $\sim 13$~s is hard to explain, because modes with $\ell= 1$ should have higher observable amplitudes when compared with modes to $\ell= 2$ due to geometric cancellation effects \citep{1977AcA....27..203D}. However, 
it might be possible that some $\ell= 1$ modes, and the associated period spacing, are not driven to observable amplitudes for some reason; for instance,  it might be that $\ell= 1$ modes are not excited at the effective temperature and gravity of this star but $\ell= 2$ modes are unstable. 

The other possibility in the interpretation of the period spacing of $\sim 13$~s  of  TIC~095332541 is that it is due to $\ell= 1$ modes. In this case, however,
the stellar mass should be unusually large for the standards of PG~1159 stars 
($\gtrsim 1 M_{\odot}$; see upper panel of Fig. \ref{fig:period_spacing_teff}), 
and in serious conflict with the spectroscopic mass ($\sim 0.58 M_{\odot}$), or mode trapping is severe.

We conclude that the period 
spacing $\Delta \Pi= 12.71$ s is most probably associated to quadrupole modes and that the stellar mass
derived on the basis of this period spacing is $M_{\star}= 0.55-0.57 M_{\odot}$. This 
constraint in the value of $\ell$ of the four periods at 352.567 s, 404.390 s, 416.389 s, and 441.555 s will be used in the next section when performing period fits to TIC~095332541.

\subsection{Period-to-period fits for TIC~095332541}
\label{period-to-period-fits-TIC095}

We have also performed period fits for the new GW Vir star TIC~095332541. We started
by examining the case in which we consider the 7 detected periods of this star according to 
Table \ref{table:095332541}. Unfortunately, our period-to-period fits for this star, either considering all modes as $\ell = 1$, as $\ell = 2$, or as a mixture of $\ell = 1$ and $\ell = 2$,  
do not allow us to find a clear seismological solution, that is, a single minima of 
the function $\chi^2$ distinguishable from a crowd of similar minima associated to different effective temperatures and stellar masses. Therefore, we cannot find an asteroseismological model for this star in this case.

\begin{figure} 
\includegraphics[clip,width= 1.0\columnwidth]{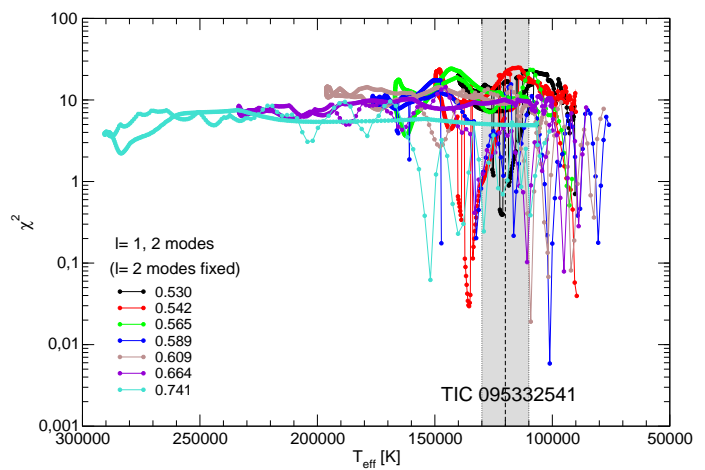}
\caption{The quality  function  of the  period  fits in terms of  the
  effective temperature  for the PG~1159 sequences with different
  stellar masses  (in solar units)  corresponding to the case in which 4
  observed periods are assumed to be associated to quadrupole ($\ell = 2$) and the remainder 2 periods can be associated either to dipole ($\ell= 1$) or quadrupole ($\ell= 2$) modes. No clear and distinguishable seismological solution is evident for TIC~095332541 in this figure.}
\label{fig:chi2-TIC095332541-l1l2} 
\end{figure} 

We also carried out period fits for this star 
following the results obtained in Sect. \ref{mean_period_spacing-TIC095}. 
That is, we assumed that a subset of 4 periods are associated with $\ell= 2$ modes, 
and left the assignment as $\ell= 1$ or $\ell= 2$ free to the remaining two periods.
Specifically, we set the "observed" value $\ell= 2$ for the periods at 352.567 s, 404.390 s, 
416.389 s, and 441.555 s, and we assume that the periods at 410.298 s and 447.293 s can be 
either $\ell= 1$ or $\ell= 2$.  The quality function versus the effective temperature for the different stellar masses is depicted in  Fig. \ref{fig:chi2-TIC095332541-l1l2}. Unfortunately, 
there is no single and clear solution in the range of effective temperature of the star, 
so, neither in this case we are able to choose any asteroseismological model for TIC~095332541.

Since we have not available an asteroseismological model for TIC~095332541, we lack the 
luminosity of the star, and that prevents us from assessing its seismological distance. More observations are needed.

\section{Summary and conclusions}
\label{conclusions}

In this paper, we have presented the discovery of two new GW Vir stars, TIC\,333432673 and TIC\,095332541. We have derived atmospheric parameters for both stars by fitting synthetic spectra to the newly obtained low to intermediate resolution SOAR/GOODMAN for TIC\,333432673  and INT/IDS for TIC\,095332541 spectra. The determined atmospheric parameters show that TIC\,333432673 and TIC\,095332541 are identical in terms of surface temperature and surface gravity ($T_{\rm eff} = 120,000 \pm 10,000$\,K and 
$\log g = 7.5\pm0.5$) and they are only different regarding the surface 
C and He abundance. 

We  investigate  the  potential  variability  of  the  two  new PG\,1159  stars  by  examining  their short and ultra-short-cadence single-sector observations obtained with \textit{TESS}.  Our frequency analysis reveals six significant independent oscillation frequencies for TIC\,333432673 and seven for TIC\,095332541, which we associate with $g$-modes.
The periodicities detected in these two PG\,1159  stars are compatible with the period spectrum typically exhibited by GW Vir stars.
TIC\,333432673 exhibits a possible contamination by other sources in the \textit{TESS} aperture within a magnitude limit of $\Delta m = 2$. We examined the target pixel files of TIC\,333432673 in order to verify the frequencies that are originated by the PG\,1159. We found that the location of the 4 frequencies (S/N $\geq$ 5) is most consistent with the position of TIC\,333432673. TIC\,333432673 is a great target for ground-based time-series photometry in order to confirm \textit{TESS} signals. 
Furthermore, TIC\,333432673 and TIC\,095332541 are excellent targets for the future ground-based photometric monitoring in order to possibly measure rates of period change and compare it to the theoretical predictions.

We have carried out an asteroseismological investigation on both GW Vir stars 
employing fully evolutionary models of PG~1159 stars. For TIC~095332541 we have 
been able to find a possible constant period spacing, and therefore it was possible 
to estimate a stellar mass in the range $0.55-0.57 M_{\odot}$ based on that quantity. 
Unfortunately,  we did not find an asteroseismological 
model for this star, which prevented us from estimating its structural parameters and its seismological distance. For TIC~333432673 we have been able 
to find a very clear constant period spacing that leads to a stellar mass in 
very good agreement with the spectroscopic mass. Also for this star, we were able to find an asteroseismological model whose mass is in excellent agreement with the 
spectroscopic mass, and the derived seismological distance is in concordance with 
the distance estimated by {\it Gaia} for this star.
 
In this paper, we have discovered and characterized two new GW Vir stars using the high-quality data collected by the {\it TESS} space mission and follow-up spectroscopy. So far, the TESS mission has met the expectations of researchers both in the goal of finding new planetary systems, and in terms of new findings in the area of stellar seismology. We are confident that this mission will continue to provide exciting new discoveries in the field of pulsating WD and pre-WD asteroseismology.

\begin{acknowledgements}
 We  wish  to  acknowledge  the  suggestions  and comments of an anonymous referee that strongly improved the original version of this work. 
 M.U. acknowledges financial support from CONICYT Doctorado Nacional in the form of grant number No: 21190886 and ESO studentship program.
 SOK is supported by CNPq-Brazil, CAPES-Brazil and
 FAPERGS-Brazil. K.J.B. is supported by the National Science Foundation under Award AST-1903828.
 This  paper includes data collected with the TESS mission, obtained
 from the MAST data archive at the Space Telescope Science Institute
 (STScI). Funding for the TESS mission is provided by the NASA Explorer
 Program. STScI is operated by the Association of Universities for
 Research in Astronomy, Inc., under NASA con-tract NAS 5–26555.  Part
 of this work was supported by AGENCIA through the Programa de
 Modernizaci\'on Tecnol\'ogica BID 1728/OC-AR, and by the PIP
 112-200801-00940 grant from CONICET. 
 This research has made use of NASA's Astrophysics Data System.
 Financial support from the National Science Centre under projects No.\,UMO-2017/26/E/ST9/00703 and UMO-2017/25/B/ST9/02218 is appreciated.
\end{acknowledgements}

\bibliographystyle{aa}
\bibliography{paper_bibliografia.bib}

\end{document}